\documentclass[pss]{wiley2sp} % provides pss two-column style
\usepackage{amsmath}
\begin{document}

% Title of the article
\title{On-demand entanglement generation using dynamic single-electron sources}

% Abbreviated title for the page headers
\titlerunning{On-demand entanglement}

% Authors
\author{%
  Patrick P. Hofer\textsuperscript{\Ast,\textsf{\bfseries 1}},
  David Dasenbrook\textsuperscript{\textsf{\bfseries 1}}, and
  Christian Flindt\textsuperscript{\textsf{\bfseries 2}}}

% Abbreviated list of authors for the page headers
\authorrunning{P. P. Hofer et al.}

%E-mail-address of corresponding author
\mail{e-mail
  \textsf{patrick.hofer@unige.ch}}

% author's affiliations/addresses
\institute{%
  \textsuperscript{1}\,D\'epartement de Physique Th\'eorique, Universit\'e de Gen\`eve, CH-1211 Gen\`eve, Switzerland\\
  \textsuperscript{2}\,Department  of  Applied  Physics,  Aalto  University,  00076  Aalto,  Finland}

% do not change, will be filled in by the publisher
 % do not change, will be filled in by the publisher

% Please select about four verbal keywords for your manuscript.

\abstract{%
% This is a macro for the typesetting of two-column text in an
% abstract. It will typeset the two arguments in \abstcol{}{} as the
% left and right column inside the abstract box. At the
% columnbreak there will be always a columnbreak (\par), so both
% columns start with a new paragraph. No automatic column height
% balancing is done.
%
% If used with a \titlefigure it will silently output both
% parameters as consecutive paragraphs.
%
% The macro is defined exclusively inside the argument of \abstract{};
% if used outside it will raise an error.
%
% Usage: \abstcol{<left column>}{<right column>}

We review our recent proposals for the on-demand generation of entangled few-electron states using dynamic single-electron sources. The generation of entanglement can be traced back to the single-electron entanglement produced by quantum point contacts acting as electronic beam splitters. The coherent partitioning of a single electron leads to entanglement between the two outgoing arms of the quantum point contact. We describe our various approaches for generating and certifying entanglement in dynamic electronic conductors and we quantify the influence of detrimental effects such as finite electronic temperatures and other dephasing mechanisms. The prospects for future experiments are discussed and possible avenues for further developments are identified.

  }

% The class file requires the standard graphicx Latex package. See the 'LaTeX
% standard graphics and color packages documentation' for more information at
% <http://tug.ctan.org/tex-archive/macros/latex/required/graphics/grfguide.pdf>.
%
% Accepted figure file formats depend on which LaTeX flavour is used.
% Classic LaTeX is always able to use Encapsulated Postscript (EPS);
% PDFLaTeX can't use this but accepts PDF, JPG, PNG, and GIF formats.
%
% See examples for implementing graphics in floating figure environments later in this file.
% If \titlefigure is given, it takes as its mandatory parameter the
% name (without extension) of some figure file.
\titlefigure[]{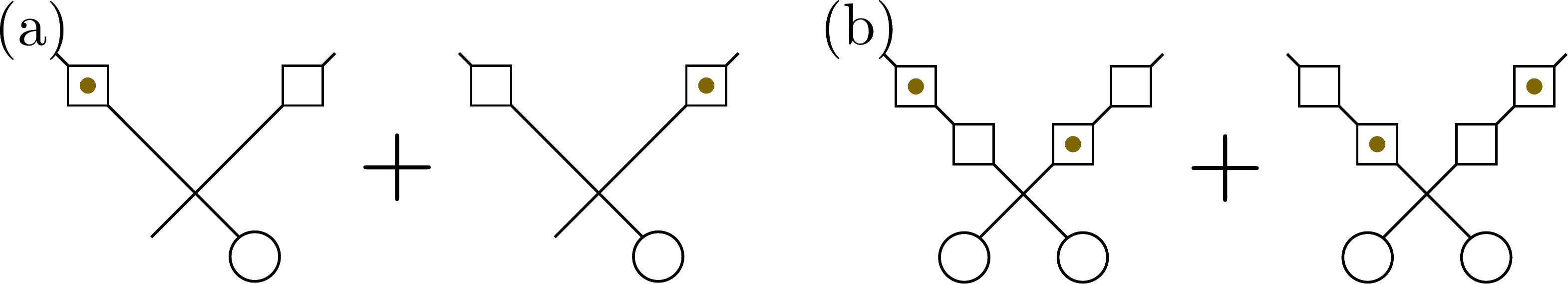}
\titlefigurecaption{%
  (a) The coherent partitioning of a single electron on a quantum point contact leads to entanglement between the outgoing arms. The entanglement can be detected using two copies of the state. (b) A time-bin entangled state is generated by partitioning two electrons on a quantum point contact followed by projection onto the subspace with one electron in each arm. The two-electron entanglement is due to the entanglement of the individual single-electron states. In both panels, circles represent single-electron sources and squares represent detectors.}

\maketitle   % please do not remove

\section{Introduction}
Recent groundbreaking experiments have demonstrated the controlled emission and interference of single-electron excitations in mesoscopic structures \cite{feve:2007,dubois:2013,bocquillon:2013,jullien:2014}. The central components of these experiments are single-electron sources combined with edge states in topological insulators (including the well-known quantum Hall effect) functioning as wave guides for the emitted electrons. Single-electron excitations can be generated using either a mesoscopic capacitor \cite{feve:2007,moskalets:2008} or by applying Lorentzian voltage pulses to an Ohmic contact \cite{dubois:2013,keeling:2006}. On-chip experiments with individual electrons can then be performed, similar to those realized with single photons on optical tables. Consequently, this field of giga-hertz quantum electronics is known as \textit{electron quantum optics} \cite{bocquillon:review}.

The controlled emission of single-electron excitations constitutes an important step towards the on-demand generation and detection of entangled single- and few electron states in mesoscopic structures. Entanglement in mesoscopic conductors with constant voltages has been investigated both experimentally and theoretically (for a review, see Ref.~\cite{beenakker:2006}). By contrast, the on-demand generation of entangled few-electron states using dynamic single-electron sources is much less explored \cite{lebedev:2005,splettstoesser:2009,sherkunov:2012,hofer:2013,inhofer:2013,vyshnevyy:2013,dasenbrook:2015,strom:2015,dasenbrook:2016,dolcetto:arxiv}. However, with the recent developments in electron quantum optics, this line of research seems promising for realizing future solid-state architectures with synchronized manipulation of quantum information encoded in electronic excitations.

In this short review, we discuss our recent proposals for generating and detecting entanglement in dynamic mesoscopic conductors. Specifically, we describe schemes for producing and detecting time-bin entanglement of electrons emitted from mesoscopic capacitors and we review our proposal for generating and detecting electron-hole entanglement in a dynamic Mach-Zehnder interferometer. Importantly, in these proposals the generated entanglement can be traced back to the partitioning of a single electron on a quantum point contact (QPC). The coherent partitioning of a single electron leads to entanglement between the outgoing arms of the QPC. Due to parity and charge super-selection rules, it is not clear how to detect the entanglement. However, together with the Quantum Correlations Group in Geneva, we have recently devised an experimental setup for the detection of single-electron (mode) entanglement. We start by describing the concept of single-electron entanglement together with the experimental scheme for its detection. We then go on to review the proposals for time-bin and electron-hole entanglement in electronic interferometers. As we show, the unifying principle of these proposals is the single-electron entanglement produced at a QPC.

The review is organized as follows. In Sec.~\ref{sec:entanglement} we discuss the basic principles of fermionic single-particle entanglement. In Sec.~\ref{sec:single-electron} we describe a scheme for detecting single-electron entanglement in an electronic Hanbury--Brown--Twiss interferometer. Sections~\ref{sec:time-bin} and~\ref{sec:electron-hole} concern our proposals for generating and detecting time-bin and electron-hole entanglement in dynamic mesoscopic structures. Finally, in Sec.~\ref{sec:conclusions} we provide our conclusions and an outlook on possible avenues for further developments.

\section{Dynamic entanglement generation}
\label{sec:entanglement}

Using a QPC, an electron can be partitioned between two separated parties named Alice and Bob, leading to the state
\begin{equation}
\label{eq:superposition}
\begin{aligned}
|\Psi_1\rangle &= \frac{1}{\sqrt{2}}\big(\hat{c}_A^\dagger+\hat{c}_B^\dag\big)|0\rangle\\&=\frac{1}{\sqrt{2}}\big(|1_A,0_B\rangle+|0_A,1_B\rangle\big),
\end{aligned}
\end{equation}
where $\hat{c}_\alpha^\dagger$ creates a particle localized with party $\alpha=A,B$ and $|0\rangle=|0_A\rangle\otimes|0_B\rangle$ denotes the state with no (excess) particles with either party. In the second line, we indicate the number of particles localized with each party, i.~e.~$|1_A,0_B\rangle=|1_A\rangle\otimes|0_B\rangle$ and $|0_A,1_B\rangle=|0_A\rangle\otimes|1_B\rangle$. Together with the Quantum Correlations Group in Geneva, we have recently demonstrated that this state is entangled and the entanglement can be experimentally detected \cite{dasenbrook:2016,friis:2016perspective}. At first, this might not be obvious, since the local operations that Alice and Bob can perform are constrained by parity and charge super-selection rules. Alice and Bob can only measure in the Fock basis which excludes measurements of superpositions like $|0_\alpha\rangle +|1_\alpha\rangle$ as required to violate a Bell inequality.

The entanglement in Eq.~\eqref{eq:superposition} can nevertheless be detected using a copy of the state \cite{bartlett:2007,dasenbrook:2016}. Since two fermions cannot occupy the same state, we introduce an additional degree of freedom that we denote by $\sigma=\pm$. It could be the spin of the particles \cite{lebedev:2004}, but could equally well be any other degree of freedom such as the nature of the particles (electron or hole) \cite{dasenbrook:2015} or the times that the particles traverse the QPC \cite{splettstoesser:2009,hofer:2013}. The state obtained by taking two copies of Eq.~\eqref{eq:superposition} then reads
\begin{equation}
\label{eq:twopstate}
\begin{split}
|\Psi_2\rangle &= \frac{1}{2}\big(\hat{c}_{A,+}^\dagger+\hat{c}_{B,+}^\dagger\big)\otimes\big(\hat{c}_{A,-}^\dagger+\hat{c}_{B,-}^\dagger\big)|0\rangle\\
&=\frac{1}{2}\big(|A_+\rangle+|B_+\rangle\big)\otimes\big(|A_-\rangle+|B_-\rangle\big),
\end{split}
\end{equation}
where $|A_\sigma\rangle=\hat{c}^\dagger_{A,\sigma}|0\rangle$ and $|B_\sigma\rangle=\hat{c}^\dagger_{B,\sigma}|0\rangle$. Here, we have partitioned the two-particle Hilbert space $\mathcal{H}$ according to the additional degree of freedom as
\begin{equation}
\label{eq:pmpartition}
\mathcal{H}=\mathcal{H}_+\otimes\mathcal{H}_-,
\end{equation}
and the state $|\Psi_2\rangle$ is then clearly separable according to
Eq.~(\ref{eq:twopstate}). However, we may also partition the Hilbert space with respect to Alice and Bob as
\begin{equation}
\label{eq:abpartition}
\mathcal{H}=\mathcal{H}_A\otimes\mathcal{H}_B.
\end{equation}
This is a local bi-partition, since Alice and Bob are spatially separated from each other. Each subspace of this bi-partition can be spanned by the states
\begin{equation}
\label{eq:basisab}
\begin{aligned}
&|0_\alpha\rangle,\hspace{2.33cm}|2_\alpha\rangle=\hat{c}^\dagger_{\alpha,+}\hat{c}^\dagger_{\alpha,-}|0_\alpha\rangle,\\& |
+_\alpha\rangle=\hat{c}^\dagger_{\alpha,+}|0_\alpha\rangle,\hspace{.5cm} |-_\alpha\rangle=\hat{c}^\dagger_{\alpha,-}|0_\alpha\rangle,
\end{aligned}
\end{equation}
with $\alpha=A,B$.
With respect to this bi-partition, the two-particle state reads
\begin{equation}
\label{eq:twopstateab}
|\Psi_2\rangle = \frac{1}{2}\left(|2_A,0_B\rangle+|0_A,2_B\rangle+|+_A,-_B\rangle+|-_A,+_B\rangle\right)
\end{equation}
with $|2_A,0_B\rangle=|2_A\rangle\otimes|0_B\rangle$ and so forth. This state cannot be written as a product of a state belonging to $\mathcal{H}_A$ and a state belonging to $\mathcal{H}_B$. Thus, we see that $|\Psi_2\rangle$ is separable with respect to the $\pm$-partition, but \textit{not} with respect to the $AB$-partition. Furthermore, as we discuss in the next section, the state can be used to violate a Bell inequality without the use of measurements that are forbidden by superselection rules. This allows us to complete our argument showing that Eq.~\eqref{eq:superposition} indeed is entangled: If the state in Eq.~\eqref{eq:superposition} was not entangled, two copies of the state, as described by $|\Psi_2\rangle$, would also not be entangled. Therefore, a violation of a Bell inequality using $|\Psi_2\rangle$, demonstrating that $|\Psi_2\rangle$ is entangled, necessarily implies that $|\Psi_1\rangle$  is also entangled. Here, as well as in Ref.~\cite{dasenbrook:2016}, we consider the entanglement of formation $\mathcal{E}_F$ which is sub-additive, i.e. $\mathcal{E}_F[\hat{\rho}_1\otimes\hat{\rho_2}]\leq\mathcal{E}_F[\hat{\rho}_1]+\mathcal{E}_F[\hat{\rho}_2]$ \cite{wootters:2001}. Other measures of entanglement can lead to different conclusions \cite{wiseman:2003,wiseman:2011}.

The entanglement in $|\Psi_2\rangle$ originates from the single-particle entanglement in $|\Psi_1\rangle$. However, Alice and Bob can also use the state $|\Psi_2\rangle$  to entangle the two particles, i.~e.~create entanglement in the $\pm$-partition. Using a detector which only measures the number of particles but not their internal degree of freedom encoded in $\sigma=\pm$, Alice and Bob can project the state $|\Psi_2\rangle$ onto the subspace where each party receives a single particle. This yields the Bell state
\begin{equation}
\label{eq:stateb}
\begin{split}
|\Phi\rangle=&\frac{1}{\sqrt{2}}\big(|+_A,-_B\rangle+|-_A,+_B\rangle\big)\\
=&\frac{1}{\sqrt{2}}\big(|A_+,B_-\rangle+|B_+,A_-\rangle\big),
\end{split}
\end{equation}
where we have emphasized that the state is fully entangled with respect to both the $AB$ and the $\pm$-partition. In the $\pm$-partition, it is the projection (a local operation with respect to the $AB$-partition) which creates the entanglement \cite{wiseman:2003,vaccaro:2003,lebedev:2004}. Since the projection constitutes a non-local operation with respect to the $\pm$-partition, it is not surprising that it can create entanglement between the two particles. This is reminiscent of the entanglement distillation in Ref.~\cite{popescu:1995}.

We conclude that the entanglement between the parties Alice and Bob originates from the single-particle entanglement of the state in Eq.~\eqref{eq:superposition}. On the other hand, entanglement between the two particles can be created by projecting onto the subspace with a single particle at each location. In the following, we illustrate these concepts by reviewing our recent works on on-demand entanglement generation.

\begin{figure*}[h!]
  \centering
  \includegraphics[width=\textwidth]{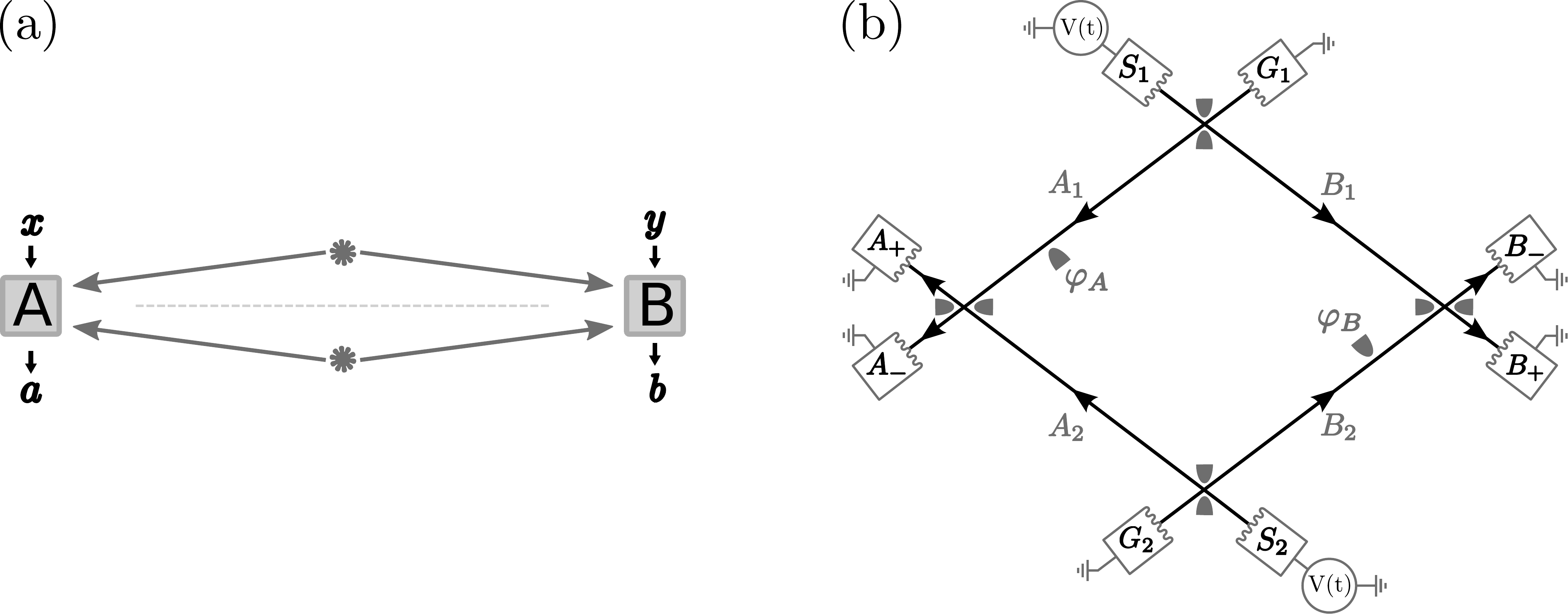}
  \caption{(a) Schematic setup for detecting single-electron entanglement. Two dynamic sources emit single electrons in superpositions towards the locations $A$ and $B$. A Bell inequality can be violated using only local operations and measurements at $A$ and $B$, indicating that the combined state of the two electrons is entangled. Since the two sources are only classically correlated through their synchronization, it follows that each single-electron source emits an entangled state. (b) Sketch of an electronic Hanbury--Brown--Twiss interferometer with single-electron excitations generated at the contacts $S_1$ and $S_2$. Local operations and measurements at the locations $A$ and $B$ enable the violation of a Bell inequality. Figure taken from Ref.~\cite{dasenbrook:2016}.}
  \label{fig:idea_setup}
\end{figure*}

\section{Single-electron entanglement}
\label{sec:single-electron}

We now explicitly show that that a single electron in a superposition of two spatially separated modes as described by Eq.~\eqref{eq:superposition} is entangled. The line of arguments follows our recent work in Ref.~\cite{dasenbrook:2016}. To access the single-electron entanglement, we make use of two copies of the state in Eq.~\eqref{eq:superposition} shared by Alice and Bob as illustrated in Fig.~\ref{fig:idea_setup} (a). As mentioned above, this allows us to circumvent the restrictions imposed by parity and charge superselection rules and it enables a violation of a Bell inequality without the use of post-selection.

We consider the electronic Hanburry-Brown-Twiss interferometer sketched in Fig.~\ref{fig:idea_setup} (b). Electrons propagate along one-dimensional chiral edge channels which are formed in the quantum Hall regime when a strong magnetic field is applied.  A setup of this type with constant voltages has previously been used to investigate the two-particle Aharonov-Bohm effect both theoretically \cite{samuelsson:2004} and experimentally \cite{neder:2007nat}. Here we consider dynamic sources $S_{1/2}$ which emit single electrons on demand, i.~e.~narrow single-electron wavepackets. The contacts $A_\pm$ and $B_\pm$ represent the detectors on Alice's and Bob's sides. The other contacts $G_{1/2}$ are grounded. We first consider a simplified situation, where the temperature is zero and we assume that single electrons above the Fermi energy can be detected one by one. A scheme relying only on measurements of the mean current and the zero-frequency noise is discussed below where we also discuss dephasing and finite temperatures.

Single-electron excitations localized in region $\alpha$ are described by the operator $\hat{A}_\alpha$ [see Eq.~\eqref{eq:levitonoperator} for an example]. The possible regions are indicated in Fig.~\ref{fig:idea_setup} (b). The electrons are emitted from a source at the time $t=0$ and reach the region $\alpha$ at the time $t_\alpha$. (We note that each region can either only be reached from a single source or it has the same distance to all possible sources due to the symmetry of the setup). For $t_\alpha=t_\beta$, the operators fulfill the canonical anti-commutation relations
\begin{equation}
\label{eq:commutation}
\{\hat{A}_\alpha,\hat{A}_\beta^\dag\}=\delta_{\alpha,\beta},\hspace{.5cm}\{\hat{A}_\alpha,\hat{A}_\beta\}=\{\hat{A}_\alpha^\dag,\hat{A}_\beta^\dag\}=0.
\end{equation}

We start by considering the case where only $S_1$ is active and produces the state $\hat{A}_{S_1}^\dag|0\rangle$ right after the source. Here, the undisturbed Fermi sea is denoted as $|0\rangle$. The electron then traverses the first QPC, and its state now reads
\begin{equation}
\label{eq:trafoqpc1}
\hat{A}_{S_1}^\dag|0\rangle=\frac{1}{\sqrt{2}}\left(\hat{A}_{A_1}^\dag+\hat{A}_{B_1}^\dag\right)|0\rangle,
\end{equation}
having assumed that the QPC is half-transparent. The equality here is due to our definition of the creation operators. Specifically, the state obtained by creating an electron in region $S_1$ at time $t_{S_1}$ is the same as the state obtained by creating an electron which is delocalized over the regions $A_1$ and $B_1$ at the later time $t_{A_1}=t_{B_1}$. We see that a single-electron source in combination with a QPC in this way can be used to prepare the state in Eq.~\eqref{eq:superposition}.

Next, the sources $S_1$ and $S_2$ which are assumed to be identical are operated synchronously. Importantly, the classical correlation provided by the synchronization is the only correlation between the sources and it cannot create any entanglement. In addition to the sources, Alice at $A$ and Bob at $B$ both have a local measurement device. The measurement device consists of a half-transparent QPC mixing the two inputs channel and two detectors shown as $A_\pm$ and $B_\pm$ in Fig.~\ref{fig:idea_setup} (b). In addition, the phases $\varphi_{A/B}$ applied at one of the inputs is tunable. Experimentally, the tunable phase can be provided by a side-gate or, since only the sum of the phases $\varphi=\varphi_A+\varphi_B$ matters, by a magnetic flux threading the interferometer.  The degree of freedom that distinguishes the two particles (called $\pm$ in the last section) is the orbital degree of
freedom given by the input in which the electron arrives to Alice ($A_1$ or $A_2$) or Bob ($B_1$ or $B_2$).

The two-electron state moving from the sources to the detectors can now be written
\begin{equation}
\begin{aligned}
\label{eq:statev}
\hat{A}_{S_1}^\dagger \hat{A}_{S_2}^\dagger &|0\rangle  = \frac{1}{2} \left(\hat{A}_{A_1}^\dagger e^{i\varphi_A} +
\hat{A}_{B_1}^\dagger\right)\left(\hat{A}_{A_2}^\dagger + \hat{A}_{B_2}^\dagger e^{i\varphi_B}\right) |0\rangle\\&  = \frac{1}{4} \bigg[ \left(\hat{A}_{A_+}^\dagger \hat{A}_{B_+}^\dagger+\hat{A}_{A_-}^\dagger \hat{A}_{B_-}^\dagger\right) \left(e^{i\varphi}-1\right) \\&\qquad+ \left(\hat{A}_{A_+}^\dagger \hat{A}_{B_-}^\dagger+\hat{A}_{A_-}^\dagger \hat{A}_{B_+}^\dagger\right) \left(e^{i\varphi}+1\right)
 \\&\qquad- 2 e^{i\varphi_A} \hat{A}_{A_+}^\dagger \hat{A}_{A_-}^\dagger + 2 e^{i\varphi_B} \hat{A}_{B_+}^\dagger
\hat{A}_{B_-}^\dagger \bigg] |0\rangle,
\end{aligned}
\end{equation}
where we assumed that the propagation along the edge channels has the same effect on both emitted electrons.
Assuming that $\hat{A}^\dag_{A_\pm}$ induces a click in Alice's detector $A_\pm$ (and similarly for Bob), a Bell inequality can be violated using the following strategy: The binary inputs $x,y=0,1$ determine the phases $\varphi_A^x$ and $\varphi_B^y$. The binary outputs $a,b=\pm1$ are then obtained by outputting $a=\pm1$ ($b=\pm1$) if detector $A_\pm$ ($B_\pm$) clicks. If both or none of the detectors click, the outputs are defined to be $+1$ and $-1$ respectively. Denoting the probability for outputs $a$, $b$ given inputs $x$, $y$ by $P(ab|xy)$, the correlator defined as
\begin{equation}
E_{xy} = \sum_{a,b} a b P(ab|xy)
\end{equation}
is then given by
\begin{equation}
 \label{eq:Exycorrelator}
E_{xy}  = -\frac{1+\cos(\varphi_A^x + \varphi_B^y)}{2} .
\end{equation}
If the experiment can be explained by a local hidden variable model, the Clauser-Horne-Shimony-Holt (CHSH) inequality should hold \cite{clauser:1969}
\begin{equation}
\label{eq:chsh}
S = | E_{00} + E_{01} + E_{10} - E_{11} | \leq 2 .
\end{equation}
However, with the choice $\varphi_A^0=0$, $\varphi_A^1=\pi/2$, $\varphi_B^0=-3\pi/4$, and
$\varphi_B^1=3\pi/4$, we find
\begin{equation}
S = 1+\sqrt{2} > 2,
\end{equation}
which is clearly violating the CHSH inequality. Since the two-electron state was obtained by local operations on single-electron states originating from independent sources, which at most can be classically correlated due to their synchronization, we conclude that the single-electron state in Eqs.~\eqref{eq:superposition} and \eqref{eq:trafoqpc1} must be entangled. The Bell inequality violation is not subject to the detection loophole as our scheme does not involve post-selection \cite{brunner:2014}.

We now turn to an experimental scheme which can be realized with current technology. The single-electron excitations can be generated by applying Lorentzian voltage pulses to an Ohmic contact \cite{keeling:2006,dubois:2013} or by using a mesoscopic capacitor \cite{feve:2007,moskalets:2008}. Here we focus on voltage pulses. The excitations created in this manner are known as \textit{levitons} and can be described using Floquet scattering theory \cite{moskalets:book}. A time-dependent voltage can be described by the phase
\begin{equation}
  \label{eq:timedepphase}
  \phi(t) = - \frac{e}{\hbar} \int_{-\infty}^t V(t') \mathrm{d} t',
\end{equation}
which electrons pick up upon leaving the contact (which is then described as being in local equilibrium). Levitons are created by applying periodic Lorentzian voltage pulses
\begin{equation}
  \label{eq:lorentzianvoltage}
  eV(t) = \sum_{j=-\infty}^{\infty} \frac{2\hbar \Gamma}{\left(t-n\mathcal{T} \right)^2 + \Gamma^2}
\end{equation}
with
\begin{equation}
\label{eq:intcharge}
\frac{e}{h}\int_{0}^{\mathcal{T}}V(t)dt=1,
\end{equation}
where $\mathcal{T}=2\pi/\Omega$ is the period and $\Gamma$ the width of the pulses. The probability amplitude for an electron to change its energy by $n$ energy quanta $\hbar\Omega$ due to the voltage reads
\begin{equation}
  \label{eq:levitonfloquetmatrix}
\begin{aligned}
S_n&=\frac{1}{\mathcal{T}}\int_{0}^{\mathcal{T}}e^{in\Omega t}e^{-i\phi(t)}\\&= \begin{cases}
    -2 e^{-n\Omega \Gamma} \sinh(\Omega \Gamma) & n > 0 \\
    e^{-\Omega \Gamma} & n=0 \\
    0 & n < 0\end{cases}.
\end{aligned}
\end{equation}
The probability to lose energy ($n<0$) is zero implying that clean electron-like excitations are created without any accompanying particle-hole pairs \cite{dubois:2013prb}. For an infinite period, these are described by the annihilation operator \cite{keeling:2006}
\begin{equation}
  \label{eq:levitonoperator}
  \hat{A}_\alpha = \sqrt{2 \Gamma} \sum_{E>0} e^{(it_\alpha-\Gamma) E/\hbar} \hat{a}_\alpha(E),
\end{equation}
where $\hat{a}_\alpha(E)$ annihilates the scattering state which looks like a plane wave of unit flux in region $\alpha$ and $t_\alpha$ denotes the time at which the levitons appear in region $\alpha$ (due to the chirality of the edge states, the direction of propagation is fixed). The Floquet scattering matrices of the full structure is  obtained by combining the amplitudes for the sources with the scattering matrices of half-transparent QPCs and the phases $\varphi_{A/B}$ (for details, see Ref.~\cite{dasenbrook:2016}).

In order to violate a Bell inequality using only measurements of the mean current and the zero-frequency noise \cite{kawabata:2001,chtchelkatchev:2002,samuelsson:2003,samuelsson:2004}, we need to assume that the long-time measurements amount to ensemble averages over the state in a single period. This is a reasonable assumption if the width of the levitons is much smaller than the period $\Gamma\Omega\ll1$. The levitons are then well-separated and correlations between electrons emitted within different periods can safely be neglected. At zero temperature, the Fermi sea does not contribute to the long-time measurements and the dc current operator can be expressed in terms of the leviton number operator as
\begin{equation}
  \label{eq:levitonnumber}
 \hat{I}_\alpha=\frac{e}{\mathcal{T}}\hat{A}_\alpha^\dagger \hat{A}_\alpha
\end{equation}
The current operators at detector $A_+$ can then be written as
\begin{equation}
\hat{I}_{A_+}=\frac{e}{2\mathcal{T}}\left(\hat{A}_{A_1}^\dag\hat{A}_{A_1}+\hat{A}_{A_2}^\dag\hat{A}_{A_2}+\hat{\sigma}^A_{\varphi_A}\right),
\end{equation}
where $\hat{\sigma}^A_{\varphi_A}=[\cos(\varphi_A)\hat{\sigma}_x^A+\sin(\varphi_A)\hat{\sigma}_y^A]$ and $\hat{\sigma}_{x/y}^A$ are the usual Pauli matrices in the basis $\hat{A}^\dag_{A_1}|0\rangle$, $\hat{A}^\dag_{A_2}|0\rangle$.
With this in mind, we define the operators
\begin{equation}
  \label{eq:Xoperators}
  \begin{aligned}
  &\hat{X}_A^{\varphi_A} = \frac{2\mathcal{T}}{e}\left( \hat{I}_{A_+} - \langle\hat{I}_{A_+}\rangle\right) ,\\&
  \hat{X}_B^{\varphi_B} = \frac{2\mathcal{T}}{e}\left( \hat{I}_{B_+} - \langle\hat{I}_{B_+}\rangle\right).
  \end{aligned}
\end{equation}
In the single-electron subspace, where Alice and Bob both receive a single electron, these operators measure the Pauli operator rotated by the controllable angle $\varphi_A$. If two or no electrons arrive at the detector, the operators will give the contributions $+1$ or $-1$ respectively (to see this, one may use that $\langle \hat{I}_{A_+}\rangle=\langle \hat{I}_{B_+}\rangle=e/{2\mathcal{T}}$). At zero temperature, these operators yield the correlators
\begin{equation}
\label{eq:currentcorr}
\langle \hat{X}_A^{\varphi_A} \hat{X}_B^{\varphi_B} \rangle = - \frac{1+\cos(\varphi_A+\varphi_B)}{2},
\end{equation}
showing that they fulfill the same statistics as the strategy discussed above Eq.~\eqref{eq:Exycorrelator}, where we assumed the possibility to detect single electrons. The CHSH inequality in Eq.~\eqref{eq:chsh} can thus be violated in an analogous way.

The average current and the zero-frequency noise do not depend on the pulse width $\Gamma$. Still, the assumption $\Gamma\Omega\ll1$ is crucial to infer single-electron entanglement as it ensures that only a single electron from each source traverses the interferometer at any given time. This is different from Ref.~\cite{samuelsson:2004}, where a constant voltage source was used and the violation of a Bell inequality was discussed in terms of two-particle orbital entanglement obtained by projecting onto the subspace where Alice and Bob each obtain one particle. Due to this projection, a maximal violation of the CHSH inequality can be achieved ($S=2\sqrt{2}$).

\begin{figure}
  \centering
  \includegraphics[width=\columnwidth]{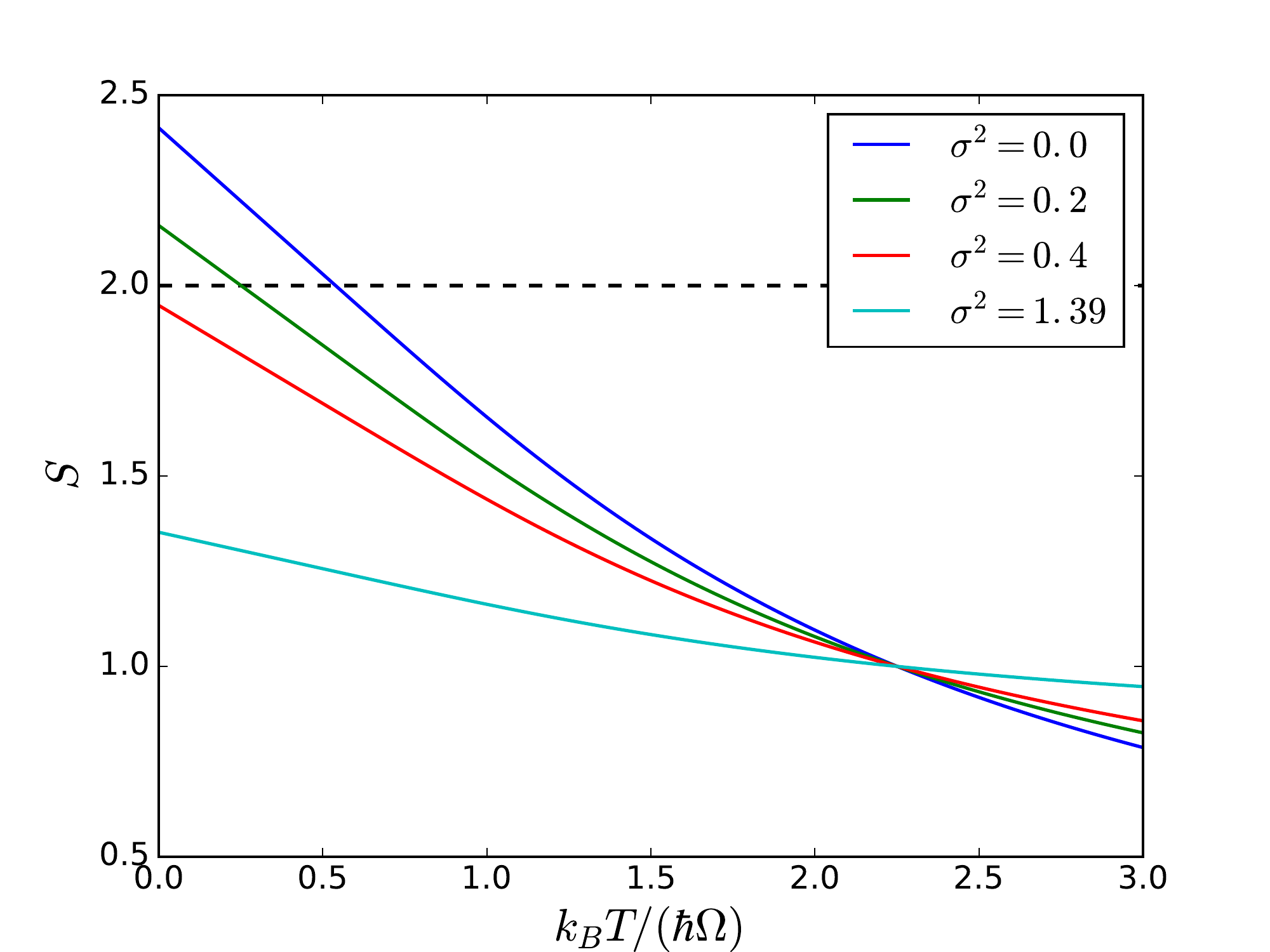}
  \caption{Maximal value of the CHSH parameter as a function of temperature. The dephasing parameter $\sigma^2$ is the variance of the distribution of the sum of the phases $\varphi_A+\varphi_B$. The dashed line indicates the CHSH bound. The figure is taken from Ref.~\cite{dasenbrook:2016}.}
  \label{fig:chshtemperature}
\end{figure}

At finite temperatures, Eq.~\eqref{eq:levitonnumber} is no longer valid and the meaning of a CHSH violation becomes more subtle. We show in Appendix~\ref{app:finitetemperatures} that entanglement can still be demonstrated for narrow pulses. Furthermore, dephasing due to interactions within the system itself or with an external environment decreases the parameter $S$. Dephasing is taken into account with a phenomenological model which describes Gaussian phase noise with variance $\sigma^2$ \cite{samuelsson:2009,dasenbrook:2016}. The effect of finite temperatures and dephasing is illustrated in Fig.~\ref{fig:chshtemperature}, showing that entanglement can still be detected under realistic experimental conditions.

Having established that the coherent splitting of a single electron on a quantum point contact produces an entangled state, we now turn to our proposals for create two-electron entanglement based on this essential principle.

\begin{figure*}
  \centering
  \includegraphics[width=.9\textwidth]{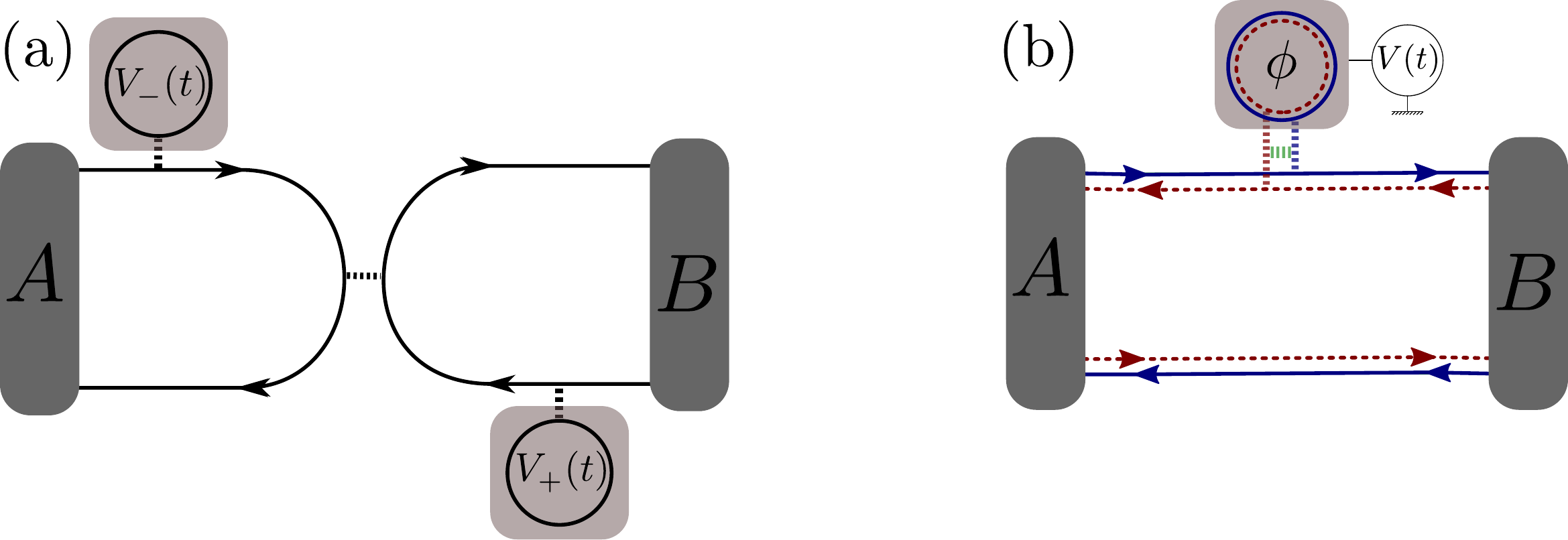}
  \caption{(a) Two mesoscopic capacitors which emit single electrons onto a common QPC with a time delay can be used to create time-bin entanglement \cite{splettstoesser:2009}. (b) Using the edge states of a quantum spin Hall insulator, a single source can be used to create time-bin entanglement \cite{hofer:2013}. The role of the central QPC in (a) is played by the QPC connecting the capacitor to the elongated edge state. The delay between electron emissions is controlled by a magnetic flux $\phi$ which locally breaks time-reversal symmetry. }
  \label{fig:twopartent}
\end{figure*}

\section{Time-bin entanglement}
\label{sec:time-bin}

We now discuss the creation of time-bin entanglement using single-electron sources and a quantum point contact \cite{splettstoesser:2009,hofer:2013}. As discussed above, an entangled two-particle state can be created using two copies of a delocalized single-electron state together with a projection onto the subspace where each party receives one electron. Here, we are interested in the case where the entangled degrees of freedom are provided by the time that the electrons arrive at their respective parties Alice and Bob. To this end, we consider a QPC with two inputs and two outputs leading to Alice and Bob. The first electron is sent onto the QPC at time $t_-$ creating the state
\begin{equation}
\label{eq:psimin}
\begin{split}
|\Psi_-\rangle &=\frac{1}{\sqrt{2}}\left(|-_A\rangle+|-_B\rangle\right)\\
&=\frac{1}{\sqrt{2}}\left(\hat{A}_{A,-}^\dag+\hat{A}_{B,-}^\dag\right)|0\rangle
\end{split}
\end{equation}
Here, the first subscript denotes the spatial localization and the second denotes the time, i.~e.~$\hat{A}_{A,-}^\dag$ creates an electron that is transmitted through the QPC towards Alice at time $t_-$. A second electron is then sent onto the QPC at time $t_+$, resulting in the combined two-electron state
\begin{equation}
\label{eq:psiminplus}
\begin{aligned}
|\Psi_\pm\rangle=\frac{1}{2}&\left(\hat{A}_{A,+}^\dag-\hat{A}_{B,+}^\dag\right)\left(\hat{A}_{A,-}^\dag+\hat{A}_{B,-}^\dag\right)|0\rangle\\ =\frac{1}{2}&\left(\hat{A}_{A,+}^\dag\hat{A}_{A,-}^\dag-\hat{A}_{B,+}^\dag\hat{A}_{B,-}^\dag\right.\\&\qquad+\left.\hat{A}_{A,+}^\dag\hat{A}_{B,-}^\dag-\hat{A}_{B,+}^\dag\hat{A}_{A,-}^\dag\right)|0\rangle.
\end{aligned}
\end{equation}
The different signs in the first line appear since the two electrons originate from different inputs.

If the states created by $\hat{A}_{\alpha,+}^\dag$ and $\hat{A}_{\alpha,-}^\dag$ can be perfectly distinguished, the last state is equivalent to the state in Eq.~\eqref{eq:twopstate}. Alice and Bob can then perform local projections to obtain the fully entangled Bell state in Eq.~\eqref{eq:stateb} with probability one half. On the other hand, if the times $t_-$ and $t_+$ are close, a finite overlap between the emitted electrons will reduce their distinguishability. We thus write
\begin{equation}
\label{eq:atildeplus}
\hat{A}_{\alpha,+}=J\hat{A}_{\alpha,-}+\sqrt{1-|J|}\hat{A}_{\alpha,\times}
\end{equation}
with the overlaps
\begin{equation}
\begin{aligned}
&\langle \times_\alpha|-_\alpha\rangle=\langle 0|\hat{A}_{\alpha,\times}\hat{A}^\dag_{\alpha,-}|0\rangle=0,\\&\langle+_\alpha|-_\alpha\rangle=\langle 0|\hat{A}_{\alpha,+}\hat{A}^\dag_{\alpha,-}|0\rangle=J.
\end{aligned}
\end{equation}
Here, the operator $\hat{A}_{\alpha,\times}$ creates the state obtained by projecting the overlap with $|-_\alpha\rangle$ away from the state $|+_\alpha\rangle$.

The state in Eq.~\eqref{eq:psiminplus} can now be expressed as
\begin{equation}
\label{eq:psiminplus2}
\begin{split}
|\Psi_\pm\rangle =&\frac{\sqrt{1-|J|^2}}{2}\left(|-_A,\!\times_B\rangle\!+\!|\times_A,\!-_B\rangle\right)\!+\!J^*|-_A,-_B\rangle\\
&+\frac{\sqrt{1-|J|^2}}{2}\left(|2_A,0_B\rangle-|0_A,2_B\rangle\right),
\end{split}
\end{equation}
where $|2_\alpha\rangle=\hat{A}^\dag_{\alpha,\times}\hat{A}^\dag_{\alpha,-}|0\rangle$ and $|i_A,j_B\rangle=\hat{A}^\dag_{A,i}\hat{A}^\dag_{B,j}|0\rangle$ with $i,j\in-,\times$. If the emitted electrons are fully overlapping ($t_{+}=t_-$ and $|J|=1$), Alice and Bob each receive one electron and the state is separable. This is the fermionic Hong-Ou-Mandel effect. With no overlap ($J=0$), we recover the state in Eq.~\eqref{eq:twopstateab} (up to a relative phase which is not important in this limit).

We now consider the situation where Alice and Bob are restricted to measurements  which are diagonal in the Fock basis. Then the entanglement between subspaces with different local particle numbers is inaccessible as discussed above \cite{beenakker:2006}. However, with a local measurement of the total particle number, Alice and Bob can project onto the subspace where each party receives one electron. The resulting state is given by the first line of Eq.~\eqref{eq:psiminplus2}. This state lives in the subspace spanned by $|-_\alpha\rangle$ and $|\times_\alpha\rangle$ with $\alpha=A,B$ which is equivalent to the Hilbert space for two-qubit states shared by Alice and Bob. This entanglement can be quantified by the concurrence which is easily evaluated for pure states \cite{wootters:1998,splettstoesser:2009,hofer:2013}. For the state after projection, it reads
\begin{equation}
\label{eq:concurrence}
\mathcal{C}=\frac{1-|J|^2}{1+|J|^2},
\end{equation}
which reaches unity (zero) for zero (complete) overlap. To obtain the entanglement generated in a single realization, the concurrence should be multiplied with the probability of obtaining one electron with each party. This gives the concurrence per cycle
\begin{equation}
\label{eq:concpercycle}
\tilde{\mathcal{C}}=\frac{1-|J|^2}{2},
\end{equation}
which reaches a maximum value of $1/2$ for zero overlap.

Next, we discuss two specific proposals for creating time-bin entanglement in this way. The first proposal was put forward by Splettstoesser, Moskalets, and B\"uttiker \cite{splettstoesser:2009} and is  sketched in Fig.~\ref{fig:twopartent} (a). It uses two mesoscopic capacitors which emit single electrons (and holes) onto a central QPC. We call this the chiral proposal. Anti-bunching of electrons in such a geometry has been studied both theoretically \cite{olkhovskaya:2008} and experimentally \cite{bocquillon:2013}. The second proposal uses a single mesoscopic capacitor coupled to the helical edge states of a quantum spin Hall insulator \cite{hofer:2013}. We call this the helical proposal. We use mesoscopic capacitors that are driven adiabatically, i.~e.~the variation of the top-gate potential is slow compared to the dwell time of the electrons inside the capacitor \cite{moskalets:2013}. In this case, the mesoscopic capacitor emits a leviton described by the operator in Eq.~\eqref{eq:levitonoperator} whenever a filled level is brought above the Fermi energy. The width of the wavepacket is determined by the speed at which the level crosses the Fermi level \cite{moskalets:book}.

Using two capacitors for the chiral proposal, the scheme for time-bin entanglement can be implemented in a straightforward manner. In the helical proposal, time-bin entanglement is obtained using the spin degree of freedom. If time-reversal symmetry is preserved, the capacitor emits pairs of single electrons. This corresponds to the case $|J|=1$ and both Alice and Bob obtain an electron at the same time. A magnetic flux can be used to locally break time-reversal symmetry and split the spin-degeneracy within the capacitor. The two electrons are then emitted with  a time delay that is controlled by the magnetic flux. Allowing for spin-flips in the emission process, each electron is emitted in a superposition of a left- and right-mover in analogy to Eq.~\eqref{eq:psiminplus}.

For both proposals, the overlap reads
\begin{equation}
|J|^2=\frac{4\Gamma_-\Gamma_+}{(\Gamma_++\Gamma_-)^2+(t_+-t_-)^2},
\end{equation}
where $\Gamma_\pm$ denotes the width of the Lorentzian wave-packets. The concurrence is given by Eq.~\eqref{eq:concurrence} provided that each electron has equal probability to end up with either Alice or Bob. For arbitrary transmission probabilities, the concurrence per cycle is of the form
\begin{equation}
\label{eq:concurrenceprop}
\tilde{\mathcal{C}}=2RD(1-|J|^2).
\end{equation}
In the chiral proposal, $R$ and $D$ are the reflection and transmission probabilities of the central QPC. In the helical proposal, $R=d_\sigma/(d+d_\sigma)$ and $D=d/(d+d_\sigma)$, where $d_\sigma$ and $d$ are the transmission probabilities of the QPC connecting the capacitor to the elongated edge states with and without spin-flip, respectively.

The source of entanglement is the delocalization of the emitted single-electron states. The delocalization leads to an uncertainty in the number of electrons that are emitted towards each party which generates electrical noise. The zero frequency noise in particular is given by the concurrence obtained per cycle as \cite{hofer:2013,dolcetto:arxiv}
\begin{equation}
\label{eq:noise}
\mathcal{P}=e^2\Omega\tilde{\mathcal{C}}/\pi,
\end{equation}
where $\Omega$ denotes the frequency of the sources. We note that this equality is only valid in the absence of decoherence. However, both proposals can be extended by a Mach-Zehnder interferometer at each side in order to violate a Bell inequality which certifies the presence of entanglement \cite{splettstoesser:2009}.

\section{Electron-hole entanglement}
\label{sec:electron-hole}

In this section, the entangled degree of freedom is the charge of the electronic excitation, i.~e.~whether the quasiparticle is electron-like or hole-like \cite{dasenbrook:2015}. For a review on electron-hole entanglement using static voltage sources, see Ref.~\cite{beenakker:2006}. An electron-like excitation can be created by applying a time-dependent voltage pulse as in Eq.~\eqref{eq:timedepphase} to an Ohmic contact. A hole-like leviton can be created by reversing the sign of the voltage. The corresponding Floquet scattering matrix is then similar to the one in Eq.~\eqref{eq:levitonfloquetmatrix}, but the amplitudes for $n>0$ and $n<0$ are switched: Electrons can only lose energy, resulting in a clean hole-like excitation without additional electron-hole pairs. The electron-like leviton state is annihilated by the operator in Eq.~\eqref{eq:levitonoperator}. The corresponding annihilation operator for a hole-like leviton reads
\begin{equation}
  \label{eq:holelevitonoperator}
  \hat{A}_{\alpha,+} = \sqrt{2 \Gamma} \sum_{E<0} e^{(\Gamma - i t_\alpha)E/\hbar} \hat{a}^\dagger_\alpha(E).
\end{equation}

Similar to before, we consider a QPC with two inputs and two output channels. Instead of sending two electrons towards the QPC at different times, we now mix an electron and a hole arriving at the same time. All of the considerations from the previous section still apply. In particular, the outgoing state is given by Eq.~\eqref{eq:psiminplus}, where the indices $+$ and $-$ now denote a hole-like or electron-like excitation. They are perfectly distinguishable by their charge (and their energy), so we always have $J=0$, and the concurrence at zero temperature is always maximal.

Instead of mixing an incoming electron and an incoming hole at the QPC, the electron-hole entangled state Eq.~\eqref{eq:psiminplus} can be created directly at the QPC: By modulating the QPC transmission $D(t)$ periodically in time, it is possible to engineer a disturbance of the Fermi sea in such a way that exactly one delocalized electron-hole pair is created \cite{sherkunov:2009,zhang:2009,sherkunov:2012,dasenbrook:2015}. To see this, we consider the time-dependent scattering matrix of the QPC,
\begin{equation}
  \label{eq:timedepsmatrix}
  S(t) = \begin{pmatrix}
    r(t) & d(t) \\
    -d(t) & r(t)
  \end{pmatrix}.
\end{equation}
We choose the transmission and reflection amplitudes as
\begin{align}
  \label{eq:timedepdr}
  d(t) &= \sin \phi(t), \nonumber\\
  r(t) &= \cos \phi(t),
\end{align}
with $\phi(t)$ given by Eq.~\eqref{eq:timedepphase}. Next, we switch to the eigenbasis of $S(t)$. Particles in the two incoming eigenchannels will then be completely reflected with reflection amplitudes $r(t) \pm i d(t) = \exp(\pm i \phi(t))$ given by the eigenvalues of $S(t)$, implying that an electron-like leviton is created in one of the eigenchannels and a hole-like leviton in the other one. The physical output channels of the QPC are equal superpositions of both eigenchannels so that the electron will be delocalized over both output channels with its state given by Eq.~\eqref{eq:psimin} and similarly for the hole.

In contrast to the time-bin entanglement considered in the previous section, the entanglement here is inherently between subspaces of different local particle numbers, as the many-body state for the electron-like excitation differs from the one of the hole-like excitation by a particle number of two. However, the two basis states are in the same parity subspace \cite{friis:2016}, so there is no fundamental principle that prevents the rotation of one state into the other. Indeed, in the presence of a superconductor, an electron can be converted into a hole through an Andreev reflection. With such a setup, the violation of a Bell inequality could thus in principle be realized. As it is experimentally challenging to induce superconductivity in quantum Hall edge states, it may however be more feasible to consider an entanglement witness based on a nonlocal measurement to circumvent the particle number superselection rule.

An entanglement witness is an observable $\hat{W}$ with
\begin{equation}
  \label{eq:witness}
  \operatorname{tr} \hat{W} \hat{\rho} \leq 0
\end{equation}
for all separable states $\hat{\rho}$ on a certain subspace $\mathcal{H}_s \subseteq \mathcal{H}$ of the Hilbert space. Compared to testing a Bell inequality, the assumptions for detecting entanglement based on a witness are stronger: It must be certain that the performed measurement indeed corresponds to the operator $\hat{W}$, and that the possible states are all within $\mathcal{H}_s$. In practice, however, these assumptions can usually be justified.

One way to define an entanglement witness in quantum Hall edge channels is to combine Alice's and Bob's modes at a QPC and thereby realizing a nonlocal measurement
\cite{burkard:2003,giovanetti:2006,giovanetti:2007}. As an advantage of this strategy, the particle number superselection rule is circumvented, since the two states are mixed at the QPC. Thus, such a witness can also be used to detect electron-hole entanglement \cite{dasenbrook:2015}.

To formulate our witness, we consider a general two-particle state incident on the QPC used for the detection,
\begin{equation}
  \label{eq:generaltwoparticle}
  |\Upsilon\rangle = \sum_{\stackrel{i,j=\pm}{\alpha,\beta=A,B}} \Upsilon^{\alpha \beta}_{ij} \hat{A}_{\alpha,i}^\dagger \hat{A}_{\beta,j}^\dagger |0\rangle.
\end{equation}
If the projection of $|\Upsilon\rangle$ onto the subspace with one particle per mode is separable, $\Upsilon^{AB}$ is a rank-one matrix. In this case, the zero-frequency cross-correlator $S_{34} = \langle \hat{I}_3 \hat{I}_4 \rangle$ after the second QPC fulfills the inequality \cite{dasenbrook:2015}
\begin{equation}
  \label{eq:fwitness}
  f \equiv S_{34} - S_0(1 - 2 D_d R_d) \leq 0,
\end{equation}
where $D_d$ and $R_d$ are the transmission and reflection probabilities of the second QPC, and $S_0$ is $S_{34}$ for $D_d = 0$.

For a mixed state decomposed into states of the form in Eq.~\eqref{eq:generaltwoparticle}, the current correlators at the outputs will be given by a weighted average over the current correlators corresponding to each pure state in the decomposition. Due to the linearity of the witness in Eq.~\eqref{eq:fwitness}, the inequality is then still valid for separable states and constitutes an entanglement witness in the sense of Eq.~\eqref{eq:witness}. The witness, however, is not optimal, since $f$ might also be negative for some entangled states. To make the witness more useful, it is important to recall that $f$ is sensitive to the phases that are applied to the individual terms in the sum in Eq.~\eqref{eq:generaltwoparticle}. For example, between the creation and the detection, the electrons might pick up a phase $\pm \phi$ depending on their pseudo-spin quantum number $+$ or $-$ (corresponding to their electron-hole degree of freedom) in one of the channels of the conductor. This constitutes a local operation and therefore cannot create entanglement. However, the witness will generally oscillate as a function of $\phi$ and thus be violated only for certain values of the phase.

\begin{figure}
\begin{center}
  \includegraphics[width=0.9\columnwidth]{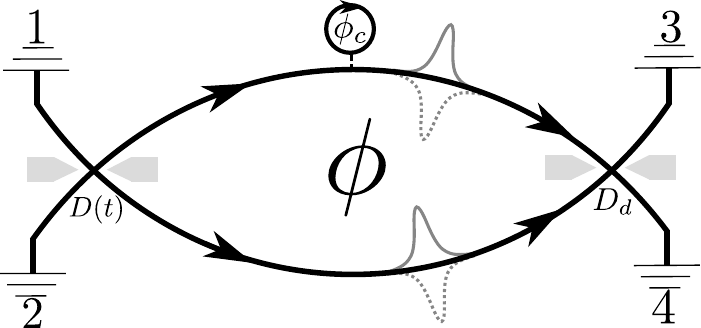}
  \caption{Setup for the creation and detection of electron-hole entanglement. Electron-hole-entangled leviton pairs are created by periodically modulating the transmission $D(t)$ of the first QPC in time. Traversing the Mach-Zehnder interferometer, the excitations pick up different phases $\phi$ (from the magnetic flux threading the device) and $\phi_c$ (from traversing a small cavity). By measuring current correlations at the outputs of the second QPC with transmission $D_d$, the electron-hole entanglement can be detected.}
  \label{fig:ehsetup}
\end{center}
\end{figure}

Combining the entanglement generation at the first QPC with the detection of the entanglement after the second QPC, the full setup takes the form of an electronic Mach-Zehnder interferometer, Fig.~\ref{fig:ehsetup}. The interferometer may be threaded by a magnetic flux $\Phi$ giving rise to a tunable phase $\phi = \pm 2 \pi \Phi / \Phi_0$ picked up by the hole-like (electron-like) levitons in the upper arm, where $\Phi_0=h/e$ is the magnetic flux quantum. Additionally, an energy-dependent phase $\vartheta(E) = \Theta(E) (\phi_c+\pi)$ may be applied to all excitations in the upper arm of the interferometer by coupling a small cavity with flux $\phi_c$ to the edge state via a QPC with a cut-off in the transmission below the Fermi level. Only electron-like excitations will then enter the small cavity and pick up this phase. We can then tune all  relative phases in the state in Eq.~\eqref{eq:psiminplus}, which now becomes
\begin{equation}
  \label{eq:psiminplusphases}
  \begin{aligned}
  |\Psi_\pm\rangle =\frac{1}{2}&\left( -e^{i\phi_c} \hat{A}_{A,+}^\dag\hat{A}_{A,-}^\dag-\hat{A}_{B,+}^\dag\hat{A}_{B,-}^\dag\right.\\&+\left.e^{i\phi}\hat{A}_{A,+}^\dag\hat{A}_{B,-}^\dag+ e^{i(\phi_c-\phi)}\hat{A}_{B,+}^\dag\hat{A}_{A,-}^\dag\right)|0\rangle.
  \end{aligned}
\end{equation}
At zero temperature and $\phi_c = 0$, the witness $f$ becomes
\begin{equation}
  \label{eq:fexact}
  f(\phi) = - \frac{e^2 \Omega}{8 \pi} \cos(2\phi),
\end{equation}
which clearly can be positive, signaling entanglement.

\begin{figure}
  \includegraphics[width=\columnwidth]{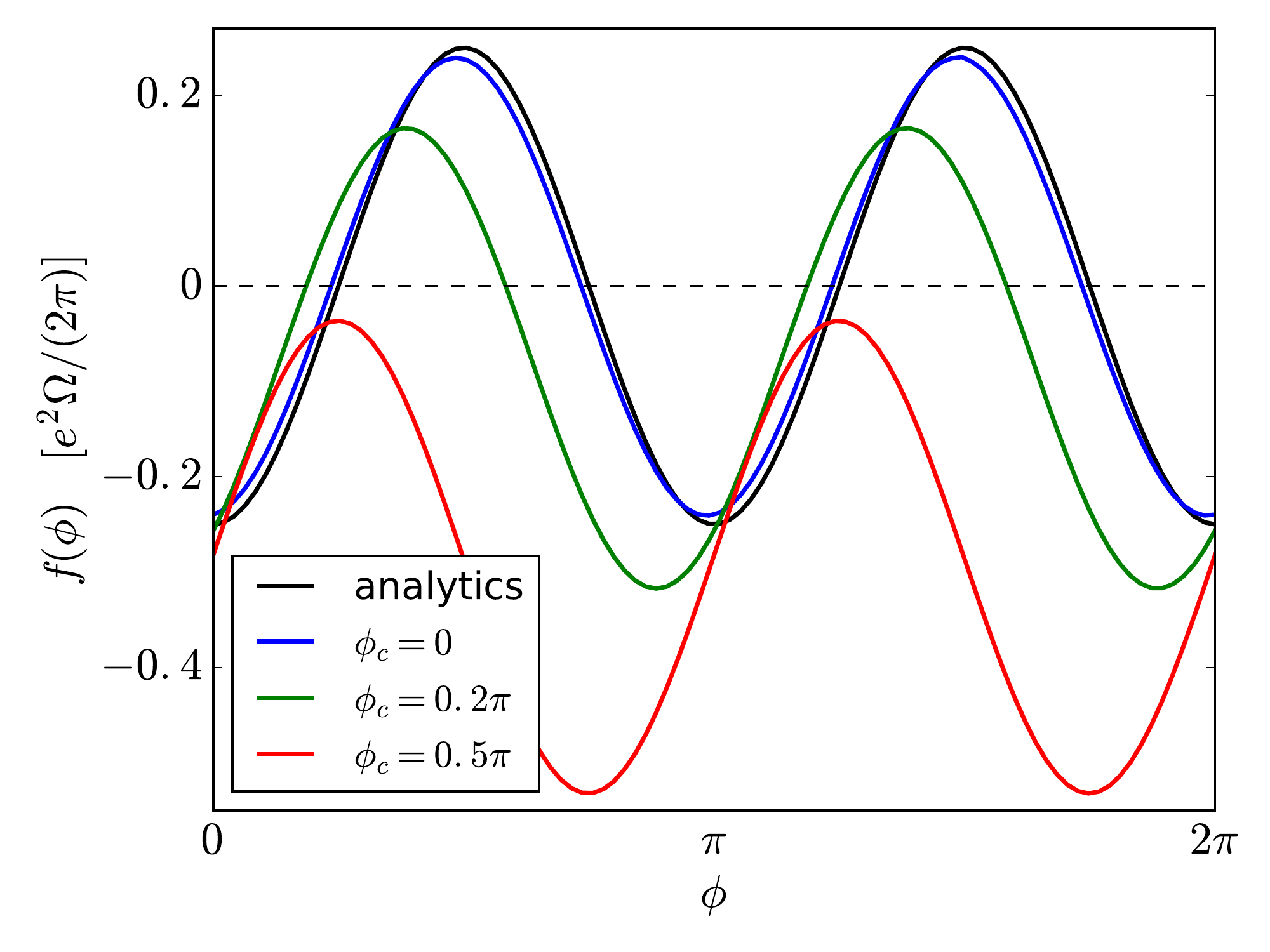}
  \caption{Entanglement witness for different values of the phases $\phi_c$ and $\phi_c$. By tuning the phases $\phi$ and $\phi_c$, the inequality \eqref{eq:fwitness} can be violated. The analytic result is given by Eq.~\eqref{eq:fexact}. The other curves are calculated using Floquet scattering theory.}
  \label{fig:fwitness}
\end{figure}

Figure~\ref{fig:fwitness} shows the witness as a function of the phases $\phi_c$ and $\phi$. For these calculations, we have used the transmission amplitude for $t_c(E) = 1 / (\exp(-\mathcal{B}E) + 1)$ for the QPC connecting the edge state to the cavity, where the cut-off $\mathcal{B}$ can be tuned by a magnetic field \cite{fertig:1987,buttiker:1990}. The zero-frequency noise is calculated using the Floquet scattering matrix approach \cite{moskalets:book}. We see that for a specific value of $\phi_c$, the system optimally violates the inequality \eqref{eq:fwitness} in the sense that the witness has the
same weight above and below the $f=0$ line as a function of $\phi$.

\begin{figure}
  \includegraphics[width=\columnwidth]{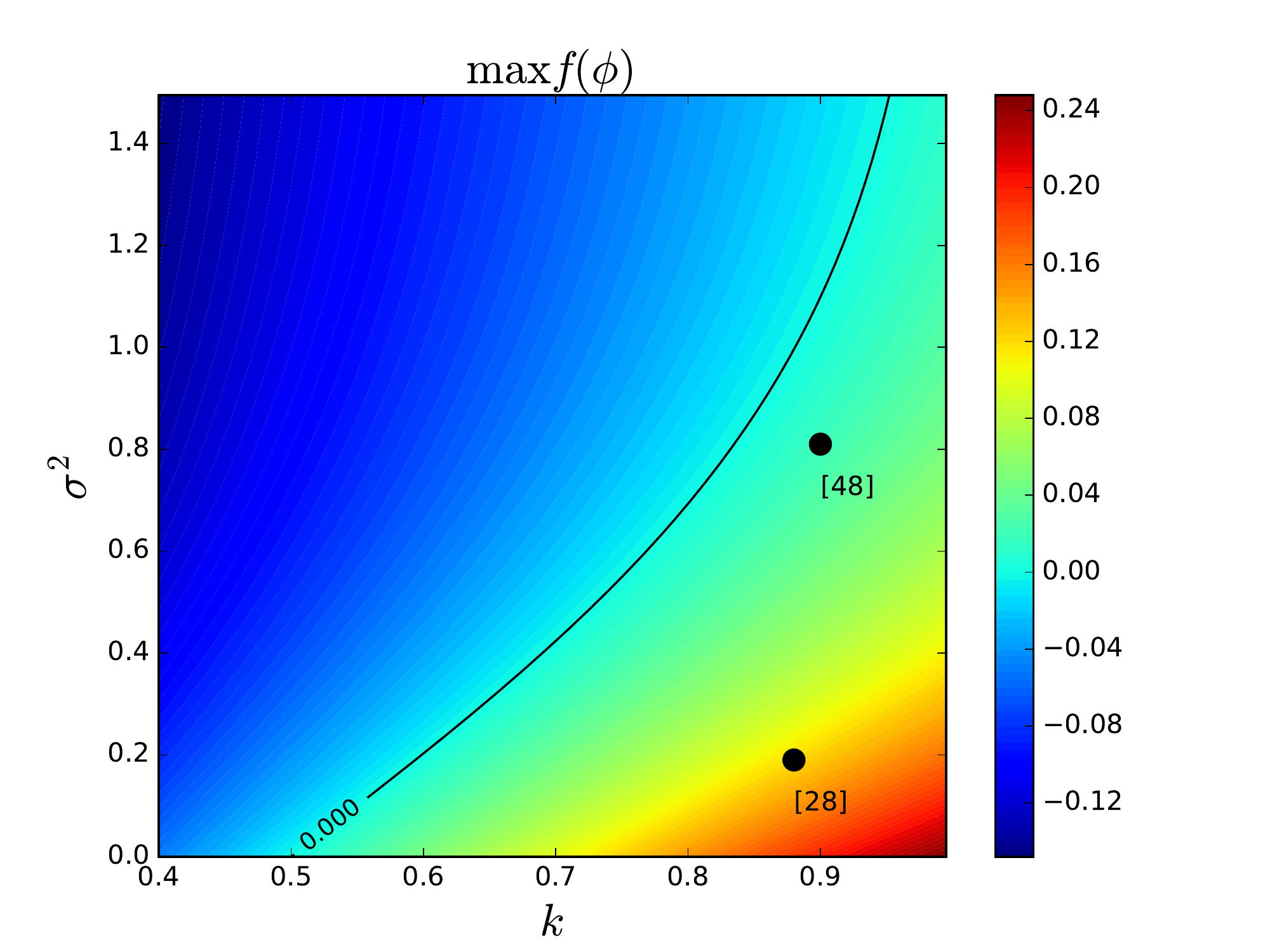}
  \caption{The maximum of the entanglement witness for different values of the phenomenological dephasing parameters $\sigma^2$ and $k$. The dots indicate the values extracted from the experiments in Refs.~\cite{neder:2007nat} and \cite{ji:2003}. For these values, the witness can be positive and the electron-hole entanglement can therefore be detected.}
  \label{fig:dephasing}
\end{figure}

The witness is more tolerant to a loss of visibility compared to entanglement certification based on
Bell inequalities which typically require a high visibility. The visibility can be lost for example
due to dephasing or a finite coherence length. Dephasing can be described by an additional
fluctuating phase $\delta \phi$ with a Gaussian distribution, picked by electrons traversing the
interferometer. The finite coherence length is characterized by the phenomenological parameter
$k$ which reduces the phase-dependent part of the transmission probability through the interferometer, see Ref.~\cite{ji:2003}. In terms of these parameters, the noise then becomes
$S_{34} = -e^2\Omega/(4\pi)[1- 2kD_dR_d \{1 - \exp(-2 \sigma^2) \cos(2\phi) \}]$. Figure~\ref{fig:dephasing} shows the maximum of the witness for different values of
$\sigma^2$ and $k$.  We see that for values corresponding to state-of-the-art experiments, it would
be possible to detect the electron-hole entanglement using this scheme. By contrast, entanglement
tests based on the CHSH inequality are still challenging \cite{neder:2007nat}.

\section{Conclusions and outlook}
\label{sec:conclusions}

We have reviewed our recent proposals for the on-demand generation and detection of few-electron entanglement in dynamic mesoscopic conductors. The common source of entanglement in these proposals is the coherent partitioning of single electrons on quantum point contacts, leading to entanglement between the two outgoing arms. To detect the single-electron entanglement, we use two copies of the state  produced by an electronic Hanbury--Brown--Twiss interferometer. The resulting two-electron state can be used to violate a Bell inequality formulated in terms of current cross-correlators measured at the output arms of the interferometer. Two dynamic single-electron emitters can also be used to generate time-bin entangled electron pairs using either chiral or helical edge states. In these cases, the zero-frequency noise detected at the outputs is a direct measure of the concurrence. Electron-hole pairs can be produced by periodically modulating the transmission of a quantum point contact. The electron-hole pairs are delocalized across the two output arms, leading to entanglement between them. By reconnecting the arms at a second quantum point, the entanglement can be detected using an entanglement witness formulated in terms of current cross-correlators.

Experimentally, the controlled generation and detection of entanglement in dynamic mesoscopic structures is in its infancy. However, several important building blocks have been experimentally realized in recent years. Dynamic single electrons sources can be operated in the giga-hertz regime, leading to measurable currents and current correlations. The emitted electrons can be guided via edge states to quantum point contacts acting as electron beam splitter. Electronic Mach-Zehnder and Hanbury-Brown-Twiss interferometers with dynamic single-electron sources may be realized in the near future. As such, the prospects for generating and detecting electronic entanglement in dynamic mesoscopic structures seem promising. We hope this review may stimulate further efforts in this direction.

\begin{acknowledgement}
We thank J. Bowles, J. B. Brask, and N. Brunner for the fruitful collaboration that led to the work on single-electron entanglement described in Ref.~\cite{dasenbrook:2016}. DD and PH gratefully acknowledge the hospitality of Aalto University. CF thanks DD and PH for visiting Aalto University. CF is affiliated with Centre for Quantum Engineering at Aalto University. The work was supported by the Swiss NSF and Academy of Finland.
\end{acknowledgement}

\appendix

\section{CHSH inequality at finite temperatures}
\label{app:finitetemperatures}
At finite temperatures, thermally populated states complicate the determination of entanglement in our system. However, a violation of the CHSH inequality using the correlators in Eq.~\eqref{eq:currentcorr} still unambiguously confirms the presence of entanglement in the many-body state in the limit $\Gamma\Omega\rightarrow0$.

To see this, we consider the reduced two-leviton density matrix defined by the relation
\begin{equation}
  \label{eq:twolevitondm}
  [\rho^{(2)}]_{\alpha \beta \gamma \delta} = \left \langle \hat{A}_\alpha^\dagger \hat{A}_\beta^\dagger
    \hat{A}_\gamma \hat{A}_\delta \right \rangle,
\end{equation}
where the expectation value is taken over the complete many-body state at a finite temperature. If it would be possible to measure quadratic and quartic correlators of the $\hat{A}$-operators at finite temperatures, the two-leviton density matrix could be reconstructed tomographically and entanglement could be verified. Experimentally, however, current correlators are typically measured. In this case, we can show that the correlators in Eq.~\eqref{eq:currentcorr} based on average currents and zero-frequency noise are bounded from above by the correlators of the leviton operators $\hat{A}$ and $\hat{A}^\dagger$. A violation of the CHSH inequality based on
current and noise measurements at finite temperatures then implies that it would also be violated based on the leviton operators, and that the two-leviton density matrix is entangled.

To this end, we calculate the correlators of the operators in Eq.~\eqref{eq:levitonoperator} directly after the sources and compare them to the measurable correlators of these operators where each $\hat{A}_\alpha^\dagger \hat{A}_\alpha$ is replaced by the zero-frequency current operator $\hat{I}_\alpha$. In order for the entanglement detection to be conclusive, we need the difference
\begin{equation}
  \label{eq:correlatordifference}
  \Delta S = S_\text{leviton number} - S_\text{current}
\end{equation}
of the CHSH parameters in Eq.~\eqref{eq:chsh} based on leviton number correlators and current
correlators to be positive. In this case, the measured correlations are always bounded from above by
the leviton number correlations. In the finite temperature case, the correlator based on the current
measurement depends on the energy ratio $k_B T / (\hbar \Omega)$, whereas the leviton correlator
depends on $\Gamma k_B T / \hbar$. In the limit $T \to 0$, the correlators equal each other.

\begin{figure}
  \centering
  \includegraphics[width=\columnwidth]{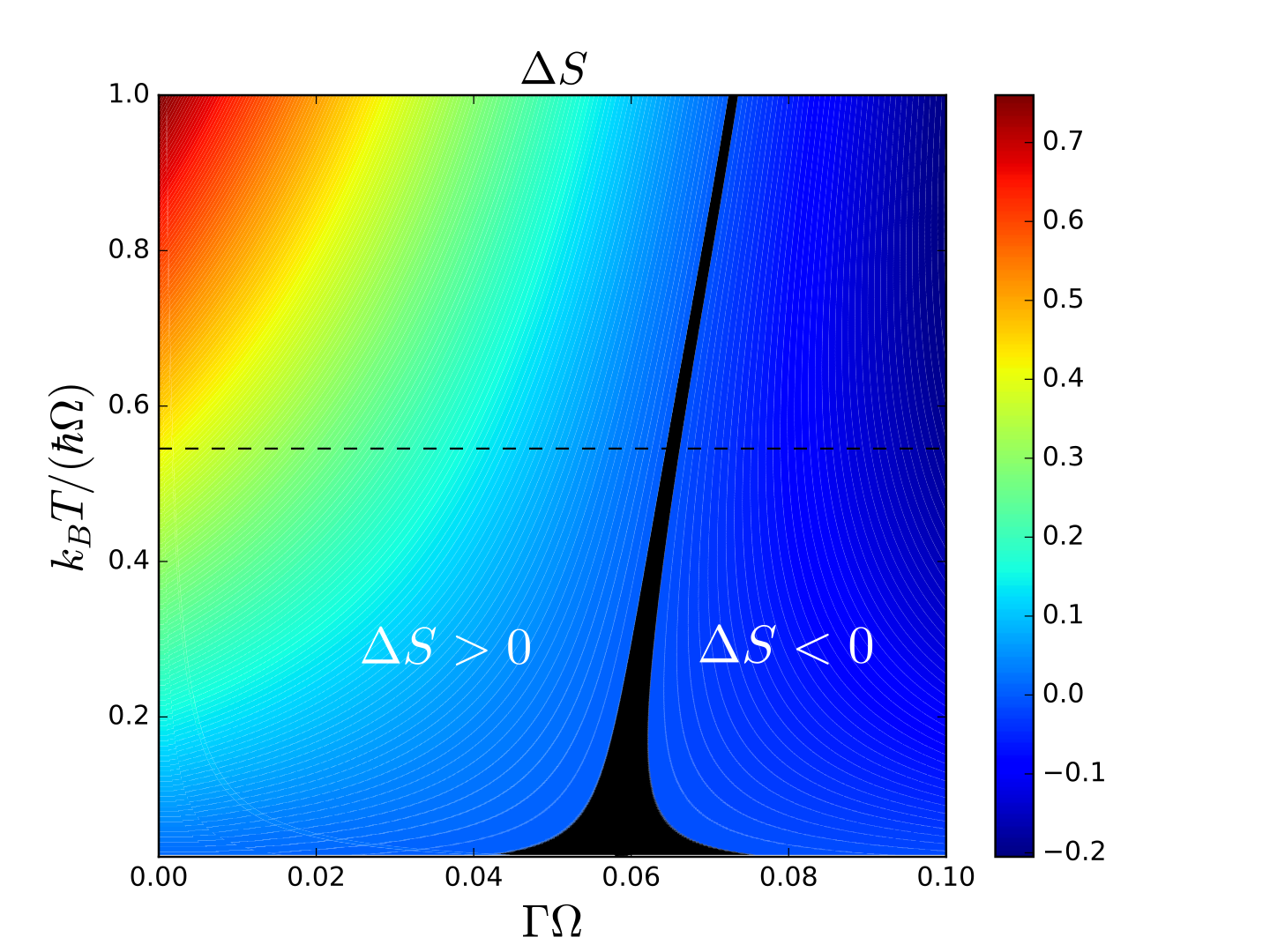}
  \caption{Difference of the CHSH parameters $\Delta S$ obtained from the leviton number correlators and from current correlators. A positive value indicates that a violation of the CHSH inequality at finite temperatures by current and noise measurements unambiguously indicates entanglement of the reduced two-leviton density matrix. The black area separates positive from negative  values. The black dashed line indicates the temperature above which the CHSH inequality can no longer be violated.}
\label{fig:chshdifference}
\end{figure}

Figure~\ref{fig:chshdifference} shows $\Delta S$ as a function of the temperature and the width of the levitons times the frequency. To detect entanglement, a low driving frequency together with sharp pulses are beneficial. This translates to a negligible overlap between the pulses or $\Gamma \Omega \ll 1$. To conclude the presence of entanglement, it is not enough to have a positive value of $\Delta S$, but the CHSH inequality also has to be violated. This is only possible for temperatures below the dashed line.

% Use the following code if you wish to generate your bibliography with BibTeX;
% replace the string "pss-demo" below with the name(s) of
% the BibTeX data base(s) you want to use.
% The resulting bibliography-output (the content of the .bbl file)
% must be pasted back into this file before submission.
% Please also include your BibTeX data base file(s) in your submission
% so that we can re-run BibTeX if necessary.
%
\bibliographystyle{pss}
\bibliography{biblio}
%
% Replace the following example bibliography with your references
% before submission:

\end{document}